\RequirePackage{fix-cm}
\documentclass[11pt]{article}
\usepackage{graphicx}
\usepackage{graphicx}% Include figure files
\usepackage{bm}% bold math
\usepackage{amssymb,amsfonts,amsmath}
\usepackage[pdftex]{color}
\usepackage{float}
\usepackage{epsfig}
\usepackage{booktabs} % for much better looking tables
\usepackage{array} % for better arrays (eg matrices) in maths
\usepackage{paralist}
\usepackage[english]{babel}
\usepackage[normalem]{ulem}
\graphicspath{{figs/}}
\usepackage{todonotes}
\usepackage{lineno}
\newcommand {\bydef}{\,\raise.07485ex\hbox{:}\kern-.025em\hbox{=}\,}
\newcommand{\0} {\textbf{0}}
\newcommand{\Lin} {\mathbb{L}\mathtt{in}}

\newcommand{\Skw}{{\mathbb{S}\textrm{kw}}}
\newcommand{\Sym}{{\mathbb{S}\textrm{ym}}}

\newcommand{\Orth} {\mathbb{O}\mathtt{rth}}

\newcommand{\dvg} {\texttt{div}\,}

\newcommand{\skw} {\mathrm{skw}\,}

\newcommand{\sym} {\mathrm{sym}\,}
\newcommand {\Bc}  {\mathcal{B}}

\newcommand {\Ec}  {\mathcal{E}}

\newcommand {\Pc}  {\mathcal{P}}

\newcommand {\Vc}  {\mathcal{V}}
\newcommand {\Wc}  {\mathcal{W}}
\newcommand {\ab} {\mathbf{a}}
\newcommand {\bb} {\mathbf{b}}

\newcommand {\eb} {\mathbf{e}}

\newcommand {\mb} {\mathbf{m}}

\newcommand {\ub} {\mathbf{u}}

\newcommand {\wb} {\mathbf{w}}
\newcommand {\zb} {\mathbf{z}}
\newcommand {\Ab} {\mathbf{A}}
\newcommand {\Bb} {\mathbf{B}}
\newcommand {\Cb} {\mathbf{C}}
\newcommand {\Db} {\mathbf{D}}
\newcommand {\Eb} {\mathbf{E}}
\newcommand {\Fb} {\mathbf{F}}
\newcommand {\Gb} {\mathbf{G}}

\newcommand {\Ib} {\mathbf{I}}
\newcommand {\Lb} {\mathbf{L}}
\newcommand {\Mb} {\mathbf{M}}

\newcommand {\Qb} {\mathbf{Q}}
\newcommand {\Rb} {\mathbf{R}}
\newcommand {\Sb} {\mathbf{S}}

\newcommand {\Yb} {\mathbf{Y}}
\newcommand {\Wb} {\mathbf{W}}
\newcommand {\Zb} {\mathbf{Z}}
\newcommand {\Do} {\mathbb{D}}

\newcommand {\Io} {\mathbb{I}}

\newcommand {\Ko} {\mathbb{K}}

\newcommand {\Mo} {\mathbb{M}}

\newcommand{\Div} {\mathrm{Div}\,}

\newcommand{\arem}{\mathbf{a}}
\newcommand{\Arem}{\mathbf{A}}
\newcommand{\aref}{\mathbf{a}_0}
\newcommand{\Aref}{\mathbf{A}_0}
\newcommand{\Omegab}{\boldsymbol\Omega}

\newcommand{\Sigmab}{\boldsymbol\Sigma}
\title{Non-affine fiber reorientation in finite inelasticity\\
}
\usepackage{authblk}
\author[1]{Jacopo Ciambella}
\author[2]{Paola Nardinocchi}

\affil[1]{Dipartimento di Ingegneria strutturale e geotecnica\\
 Sapienza Universit\`a di Roma, I-00184 Roma, Italy \\
              Tel.: 0039-06-44585293\\
              jacopo.ciambella@uniroma1.it}
\affil[2]{Dipartimento di Ingegneria strutturale e geotecnica\\
 Sapienza Universit\`a di Roma, I-00184 Roma, Italy \\
              Tel.: 0039-06-44585242\\
              paola.nardinocchi@uniroma1.it}

\begin{document}

\maketitle

\begin{abstract}
This paper introduces a model for the mechanical response of anisotropic soft materials undergoing large inelastic deformations. The composite is constituted by a soft isotropic matrix reinforced with stiff fibres, that can evolve independently from each other. 
The constitutive equations are provided in terms of the free energy density and the dissipation density which are both required to be thermodynamically consistent and structurally frame-indifferent, i.e., they must be independent of a rotation overimposed on the natural state. This is in contrast to many of the currently used inelastic models for soft fiber-reinforced materials which do not deal with the lack of uniqueness of the natural state. A constraint between the inelastic spin of the matrix and the rotation spin of the fibre is introduced to fully determined the natural state. The resulting flow rules of the inelastic processes incorporate some typical scenarios including viscoelasticity and growth.
\end{abstract}
\section{Introduction}
\label{intro}
Anisotropic soft solids are materials found either in nature and in artificial structures characterized by a soft matrix with an internal structure usually constituted by stiff fibres. Both fibres and matrix contribute to the mechanical response of the solid to actions such as forces or external stimuli like temperature, electrical, magnetic and chemical fields \cite{Sawa:2010,Ciambella:2017,Cherubini:2008}.
Modeling the inelastic behaviour of anisotropic soft solids require the formulation of evolution laws for the dissipative processes, which, in general, are induced by the inelasticity of the matrix and by the reorientation of the internal structure, if this latter can evolve independently of the matrix. Interesting examples come from biology and material science \cite{Turzi:2016,Goriely:2017,Latorre:2020,Roshanzadeh:2020,Liu:2021}.\\
Within this field, elastic and inelastic deformations are frequently described by assuming that the overall deformation $\Fb$ can be (multiplicatively) decomposed into elastic $\Fb_e$ and inelastic $\Fb_g$ parts \cite{Lee:1969}, that introduce in the modelling two layers of description. One attains to the natural state of the material, where inelastic processes take place, the other to its current state, where stresses and deformations are measured. The elastic energy and the dissipation functions are normally used to introduce suitable constitutive prescriptions compatible with thermodynamics \cite{Gurtin:2010}.\\
The lack of uniqueness of the natural state, arising from the decomposition $\Fb=\Fb_e\Fb_g$, raises several points that have been differently dealt with in the literature \cite{Nguyen:2007,Gurtin:2010,Ciambella:2021}. Moreover, when anisotropic finite inelasticity is considered, several questions remain open including the proper description of the material anisotropy in the natural state as well as the relationship between the natural state of the matrix and the one of the fibre \cite{Sansour:2003,Liu:2019}.\\
In the framework outlined above, this paper aims at addressing some of the open questions. In doing so, we propose a thermodynamically consistent model in which matrix and fibres have different natural states and analyze the consequences of the structural frame-indifference requirement. 
We start by presenting a short review of the different approaches proposed over the years to deal with the two issues cited above: the lack of uniqueness of the natural state and the description of the material anisotropy in the natural state. Then, we describe our contribution and the plan of the paper. 
\subsection{A short review}
\label{asr}
The lack of uniqueness of the natural state, originating from the multiplicative decomposition, has arisen several questions starting from \cite{Green:1971}, where the notion of structural frame-indifference was first introduced as an indifference requirement under a change of frame in the natural state, in addition to the conventional frame-indifference, i.e., a change of frame in the current configuration \cite{Gurtin:2010}. The issue is particularly significant within the framework of finite inelasticity, where the multiplicative decomposition of the deformation gradient is used to describe a wide variety of inelastic processes.\\
As highlighted in \cite{Gurtin:2010} for isotropic materials, solving the non uniqueness at a constitutive level, by requiring that the constitutive functions are structurally frame-indifferent, makes the dissipation function independent of the inelastic spin. As a consequence, the theory misses three flow rules to determine the time evolution of the inelastic component of the deformation gradient. For isotropic materials, the authors of \cite{Gurtin2005c} introduced the so-called \emph{irrotationality theorem} to show that, without lack of generality, one can set the inelastic spin to zero. For anisotropic materials, several literature contributions postulate flow rules in terms of the inelastic strain, rather than the full inelastic deformation tensor \cite{Nguyen:2007,Liu:2019}, without providing any details on the evolution of the natural state. Actually, it was shown in \cite{Ciambella:2021} that, even in the anisotropic case, the issue can be solved by assuming that the inelastic spin rate satisfies an internal constraint. With this additional equation, the theory delivers the right number of flow rules governing the time evolution of the inelastic deformation, and the dissipation function is structurally frame-indifferent.\\
In \cite{Ciambella:2021}, the problem was discussed for anisotropic solids in which the reinforcing fibers were dragged by the inelastic deformation of the matrix. However, there may be situations in which  the inelastic effects in the fibre and in the matrix are uncoupled. In \cite{Rubin:2012}, for instance, it was assumed that the inelastic part of the deformation describes the natural state of the matrix, whereas the natural state of the internal structure is described by a rotation field. This approach is indeed similar to the one proposed in this paper, but the paper \cite{Rubin:2012}, despite introducing  the definition of the kinematical framework of the theory, did not provide the evolution equations of the inelastic processes neither discussed the consequences of the structural frame-indifference requirement. A different point of view was presented in \cite{Sansour:2003,Nguyen:2007,Liu:2019}. The authors of \cite{Sansour:2003} assumed that the evolution of the internal fibre structure is driven by the inelastic part of the deformation gradient, which is recognized as a further variable of the problem and so additional evolution equations are provided. Differently, in \cite{Nguyen:2007} and \cite{Liu:2019}, it was assumed that the fibers and the matrix can exhibit a distinct time-dependent behaviour and so two different multiplicative decompositions of the deformation gradient for matrix and fiber phases were introduced. As such, the internal structure in the natural state is described by the inelastic deformation tensor  of the fiber phase. Coherently, the constitutive prescriptions involve different inelastic stretch measures and free energy densities for matrix and fiber phases, thus separate flow rules were specified.
\subsection{Our contribution}
Recently, we have studied fiber reorientation in elastic materials and considered both passive reorientation \cite{Ciambella:2019}, driven by mechanical loads, and active  reorientation \cite{Ciambella2:2019}, driven by magnetic fields. We also presented and discussed a structurally frame-indifferent model for anisotropic visco-hyperelastic materials \cite{Ciambella:2021} based on the evolution laws of the dissipative processes, which are completely determined by two scalar functions, the elastic strain energy and the dissipation densities. Therein, the anisotropic fibre structure was assumed to be dragged by elastic and inelastic deformations, without the possibility of reorienting independently of the matrix.\\
In this paper, differently from the theory presented in \cite{Ciambella:2021}, we consider an independent reorientation of the material structure and study the capabilities and the limit of the theory once thermodynamically consistency and structural frame-indifference are invoked.\\
Our approach falls within the unifying theory of material remodelling \cite{Dicarlo:2002,Rodriguez:1994} with the internal variables used to define the inelastic behaviour of the matrix and the reorientation of the fibre structure. The constitutive equations are defined in a way that they satisfy the dissipation principle and the principle of structural-frame indifference \cite{Coleman:1963,Green:1971,Gurtin:2010}, apart from the usual requirement of indifference to change of observer. It is shown that, in presence of a reorientable structure, structural frame-indifference makes the dissipation function dependent on the relative inelastic spin rate defined as the difference between the fiber reorientation spin rate and the inelastic spin rate induced by the matrix. As a consequence, the determined flow rules which are identified within the theory are not enough to solve the problem and the solution can not be uniquely determined. The issue is solved by introducing an internal constraint between the inelastic spin rate and fiber reorientation spin rate that add a further equation to be solved together with the flow rules.\\
The paper focus on transversely isotropic materials, but the theory may be straightforwardly generalized to more complex anisotropy classes. Within this class of materials, it is shown that proposed theory can describe some relevant examples from the literature, although the requirement of structural frame-indifference and the internal constraint on the spin rate limit the number of scenarios that can be encompassed.\\
Section \ref{framework} described the two-layers kinematics of the model driven by the balance equations derived in Section \ref{Beqn}. The constitutive prescriptions, both thermodynamically consistent and structurally frame-indifferent are presented and discussed in Sections  \ref{sec:SFI} and  \ref{sec:tc}. The evolution equations driving the state variables are introduced in Section \ref{ev:eqn}, whereas Section \ref{aa} present two approximations of those equations in the limit of fast or slow applied deformations.

Throughout the paper we use small bold letters to indicate vectors and capital bold letters for tensors. The inner product is indicated with a dot $\cdot$ either for vectors and tensors, i.e. $\ab\cdot\bb=\sum_{i}a_ib_i$ and $\Ab\cdot\Bb=\sum_{i,j}A_{ij}B_{ij}$, where $a_i$, $b_i$ and $A_{ij}$, $B_{ij}$ are the components. The tensor product between vectors is indicated by $\ab\otimes\bb$ and represent a tensor with components $(\ab\otimes\bb)_{ij}=a_ib_j$.
%
%
%%%%%%%%%%%%%%%%%%%%%%%%%%%%%%%%%%%%%%%%%%%%%%%%%%%%%%%%%%%%%%%%%%%%%%%%%%%%%%%%%%%
%
%
\section{The kinematical framework of the model}
\label{framework}
We identify the body with the region $\Bc$ of the Euclidean three-dimensional space $\Ec$  occupied at time instant $t=t_0$, that we denote it as \emph{reference configuration}. We introduce the vector field $\aref:\Bc \to\Vc$, with $\Vc$ the  translation space of $\Ec$, such that $\aref\cdot \aref=1$, that represents the reference orientation of the fibre at position $X$. The corresponding orientation tensor, also called \emph{structural} tensor or \emph{Finger} tensor, is given by $\Aref = \aref\otimes\aref$.\\
The deformation of the body is a time-dependent map $p :\Bc\times T\to\Ec$ that assigns at each point $X\in\Bc$ a point $x=p(X,t)$ at any instant $t$ of the time interval $T$. Accordingly, the set $\Bc_t=p(\Bc,t)$ is the configuration of the body at time $t$ and $\Bc=p(\Bc,t_0)$. We call $\ub(X,t)$ the displacement field such that $\ub(X,t)=p(X,t)-p(X,t_0)$ and we assume it to be twice continuously differentiable, such that 
\begin{equation}
\Fb(X,t)=\nabla\,p(X,t)\qquad \text{and} \qquad \dot{p}(X,t)=\dfrac{\partial p}{\partial t}(X,t)\,,
\label{DefGradient}
\end{equation}
for the deformation gradient and the referential velocity field, respectively.\\
According to the Bibly-Kr\"oner-Lee decomposition \cite{Sadik2017}, the deformation gradient \eqref{DefGradient} is decomposed into inelastic  $\Fb_g$ and elastic $\Fb_e$ tensors such that at each material point one has
\begin{equation}
\Fb(X,t)=\Fb_e(X,t)\,\Fb_g(X,t)\,.
\label{kronerlee}
\end{equation}
The inelastic deformation $\Fb_g$, is a smooth tensor-valued field with positive Jacobian determinant $J_g:=\det\Fb_g>0$, that may be the manifestation of inelastic phenomena such as growth, viscous relaxation or plasticity, and, in general, do not affect the orientation of the fibres. We remark that the relaxed (or natural) state of the matrix may not be described by a placement, meaning that $\Fb_g$ may not be the gradient of any map, or in other terms, there is no way to let each body element relaxing to its natural zero-stress state without removing the surrounding elements \cite{Rodriguez:1994,Dicarlo:2002}. Indeed, it is the elastic reversible deformation $\Fb_e$ that makes the tensor field $\Fb=\Fb_e\Fb_g$ integrable. In the following, we will call $J=\det\Fb$ and so we write $J=J_e\,J_g$ with $J_e=\det\Fb_e$.\\
We further admit the existence of a remodelling process defined by a time-dependent rotation, here identified with an orthogonal tensor ${\Rb:\Bc\times\!T \to \Orth^{+}}$, that identifies the orientation that the fibre would assume if it were freed by any force, i.e., the relaxed state of the fibre. As such, we use the notation
\begin{equation}
\Arem(X,t)=\Rb(X,t)\,\Aref(X)\Rb(X,t)^T\,,
\end{equation}
to indicate the remodeled orientation tensor, with $\Arem=\arem\otimes\arem$, and $\arem=\Rb\,\aref$ the remodelled fibre orientation. Here and henceforth, the dependence on the position $X$ and time $t$ will made explicit only when needed.\\
\begin{figure}
\begin{center}
\begin{footnotesize}
\def\svgwidth{.9\textwidth}
   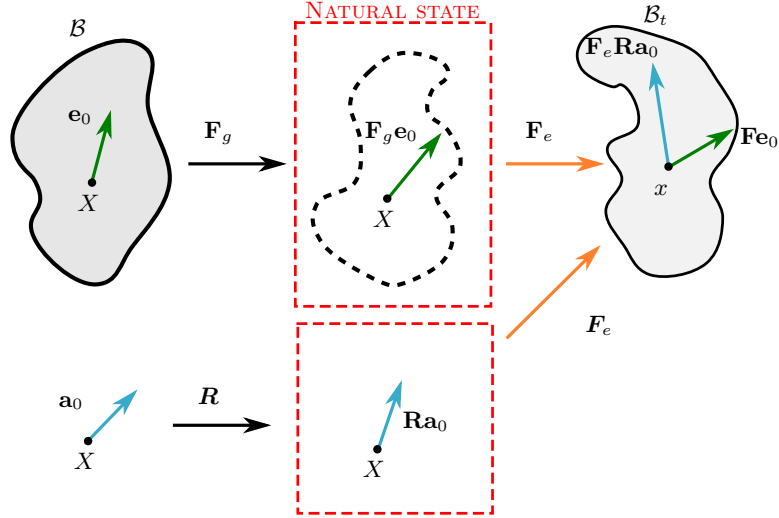
\end{footnotesize}
\end{center}
\label{fig:1}
\caption{Schematic representation of the different configurations of the matrix and the fibre in our modelling framework. The natural state is represented by a dashed line to indicate that it may not be compatible, i.e., $\Fb_g$ may not be the gradient of any map.}
\end{figure}
%
%he actual placement p, the inelastic tensor Fg and the reorientation tensor R belong to different layers of de-scription. Both Fg and R are simple kinematical entities in the relaxed state where inelastic processes belong and deliver constitutive information in the current state through the specific elastic energy and dissipation functions.
%The actual placement $p$, the inelastic tensor $\Fb_g$ and the reorientation tensor $\Rb$ represents the state variables of our model and belong to different layers of description. Both $\Fb_g$ and $\Rb$ are quantities defined in the natural state, where inelastic processes take place, and may deliver a piece of constitutive information in the current state, where strains and stresses are measured and placements  belong, through their the specific elastic energy and dissipation functions. It is worth noting that our description of the natural state differs for the matrix and the fiber, as schematically depicted in Fig.~\ref{fig:1}. Indeed, $\Fb_g$ acts on a reference line element $(X,\eb_0)$ by mapping it into the natural state as $(X,\Fb_g\eb_0)$, whereas $\Rb$ acts on a reference fiber $(X,\aref)$, producing the natural fiber state $(X,\Rb\aref)$.\\
%
The placement $p$, the inelastic tensor $\Fb_g$ and the reorientation tensor $\Rb$ represent the state variables of our model and pertain to different layers of description. Placement $p$ belongs to the current state, where strains and stresses are measured, whereas $\Fb_g$ and $\Rb$ belong to the relaxed state, where the inelastic processes take place, yet they may deliver constitutive information to the current state.
It is worth remarking that we have assumed a different relaxed state for the matrix and the fiber, as schematically depicted in Fig.~\ref{fig:1}. Indeed, $\Fb_g$ acts on a reference line element $(X,\eb_0)$ by mapping it into the natural state as $(X,\Fb_g\eb_0)$, whereas $\Rb$ acts on a reference fiber $(X,\aref)$, producing the natural fiber state $(X,\Rb\aref)$.

As largely discussed in the Literature, the multiplicative decomposition \eqref{kronerlee} causes the natural state to be non-unique. Both $\Fb_g$ and $\Qb\Fb_g$ measure the same natural state, for any $\Qb\in\Orth^+$; likewise, $\Rb$ and $\Qb\Rb$. In other words, for any $\Qb\in\Orth^{+}$ the transformations
\begin{equation}
\Fb_e^{+} = \Fb_e\Qb^T,\qquad \Fb_g^{+} = \Qb\Fb_g,\qquad \Rb^{+}=\Qb\Rb
    \label{transformation}
\end{equation}
keep the visible state unaltered, i.e.,
\begin{equation}\label{md1}
\Fb^{+}=\Fb_e^{+}\Fb_g^{+}=\Fb_e\Qb^T\Qb\Fb_g= \Fb_e\Fb_g=\Fb
\end{equation}
and
\begin{equation}\label{md2}
\ab_t^{+}=\Fb_e^{+}\Rb^{+}\aref=\Fb_e\Qb^T\Qb\Rb\aref= \ab_t\,.
\end{equation}
Several strategies have been proposed in the literature to deal with this non uniqueness including the assumption that either $\Fb_g$ or $\Fb_e$ were symmetric \cite{Nguyen:2007,Liu:2019}. A thorough discussion on this matter is presented in Sec.~\ref{sec:SFI}.

%\r{This non-uniqueness of the natural state can be dealt with by enforcing the so called principle of structural invariance (SFI) first formulated in \cite{Green:1971} that states that the constitutive prescriptions must be independent of the transformations \eqref{transformation}. The restrictions that SFI imposes on the constitutive functions will be analysed and discussed in details in Sec.~\ref{sec:SFI}.}  %Qui stiamo facendo Kinematics!!

If the triplet $(p,\Fb_g,\Rb)$ represents the local configuration space, the associated velocity triplet is $(\dot{p},\Lb_g,\Omegab)\in\Vc\times\Lin\times\Skw$ with  $\Lb_g=\dot\Fb_g\Fb_g^{-1}$ and $\Omegab=\dot\Rb\Rb^T$.

\subsection{Deformation rates}
\begin{figure}
\begin{center}
\begin{footnotesize}
\def\svgwidth{.2\textwidth}
   %% Creator: Inkscape 1.1.2 (0a00cf5339, 2022-02-04, custom), www.inkscape.org
%% Accompanies image file 'ratesfig2.pdf' (pdf, eps, ps)
\begingroup%
  \makeatletter%
  \providecommand\color[2][]{%
    \errmessage{(Inkscape) Color is used for the text in Inkscape, but the package 'color.sty' is not loaded}%
    \renewcommand\color[2][]{}%
  }%
  \providecommand\transparent[1]{%
    \errmessage{(Inkscape) Transparency is used (non-zero) for the text in Inkscape, but the package 'transparent.sty' is not loaded}%
    \renewcommand\transparent[1]{}%
  }%
  \providecommand\rotatebox[2]{#2}%
  \newcommand*\fsize{\dimexpr\f@size pt\relax}%
  \newcommand*\lineheight[1]{\fontsize{\fsize}{#1\fsize}\selectfont}%
  \ifx\svgwidth\undefined%
    \setlength{\unitlength}{141.10484086bp}%
    \ifx\svgscale\undefined%
      \relax%
    \else%
      \setlength{\unitlength}{\unitlength * \real{\svgscale}}%
    \fi%
  \else%
    \setlength{\unitlength}{\svgwidth}%
  \fi%
  \global\let\svgwidth\undefined%
  \global\let\svgscale\undefined%
  \makeatother%
  \begin{picture}(1,0.90533041)%
    \lineheight{1}%
    \setlength\tabcolsep{0pt}%
    \put(0,0){\includegraphics[width=\unitlength,page=1]{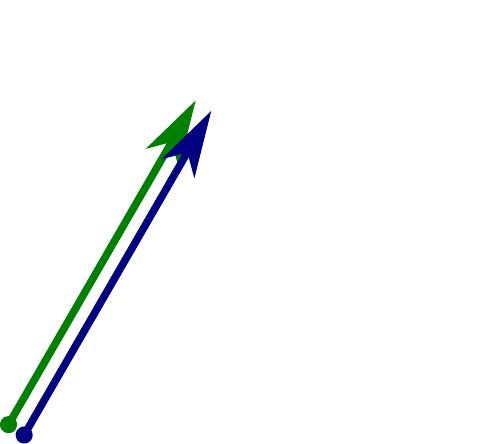}}%
    \put(0.02533191,0.45928368){\makebox(0,0)[lt]{\lineheight{1.25}\smash{\begin{tabular}[t]{l}$\mathbf{e}(\bar{t})$\end{tabular}}}}%
    \put(0.30572948,0.299871){\makebox(0,0)[lt]{\lineheight{1.25}\smash{\begin{tabular}[t]{l}$\mathbf{a}(\bar{t})$\end{tabular}}}}%
    \put(0,0){\includegraphics[width=\unitlength,page=2]{ratesfig2.pdf}}%
    \put(0.44057532,0.84132959){\makebox(0,0)[lt]{\lineheight{1.25}\smash{\begin{tabular}[t]{l}$\dot{\mathbf{e}}(\bar{t})$\end{tabular}}}}%
    \put(0.44098225,0.44301598){\makebox(0,0)[lt]{\lineheight{1.25}\smash{\begin{tabular}[t]{l}$\dot{\mathbf{a}}(\bar{t})$\end{tabular}}}}%
    \put(0.85360711,0.66337612){\makebox(0,0)[lt]{\lineheight{1.25}\smash{\begin{tabular}[t]{l}$\big(\Lb_g-\Omegab\big)\arem$\end{tabular}}}}%
  \end{picture}%
\endgroup%

\end{footnotesize}
\end{center}
\label{fig:2}
\caption{Illustration of the deformation rates for a remodelled fibre $\arem=\Rb\aref$ and a remodelled line element $\eb=\Fb_g\eb_0$, that at time $t=\bar{t}$ coincide. The difference between the two rates is given by the $\Lb_g-\Omegab$.}
\end{figure}

It is worth deriving in this section the relationships between the rates of the different kinematical quantities defined above. \\
We call $\Lb=\dot\Fb\Fb^{-1}$ the gradient of the velocity field and $\Lb_e=\dot\Fb_e\Fb_e^{-1}$ and $\Lb_g=\dot\Fb_g\Fb_g^{-1}$ the elastic and inelastic deformation rate tensors. The relationship between these quantities follows as
\[
\Lb = \Lb_e + \Fb_e\,\Lb_g\,\Fb_e^{-1}
\]
The rate of the left-Cauchy Green strain tensor $\Cb_e= \Fb_e^T\Fb_e$ is
\begin{equation}\label{rates1}
\dot\Cb_e = 2\,\Fb_e^T\Db\Fb_e - 2\,\sym\lbrace\Cb_e\Lb_v\rbrace
\end{equation}
where $\Db=\sym\lbrace \Lb\rbrace$ is the symmetric part of the velocity gradient, i.e., the stretch-rate.\footnote{Throughout the paper $\sym$ and $\skw$ will be used to indicate the symmetric and skew-symmetric part of tensors.}\\
To highlight the effects of the interaction between the matrix and the fibre, it is worth computing the rate of evolution of the remodelled fibre $\arem=\Rb\aref$ and compare it to the rate of the remodelled line element $\eb=\Fb_g\eb_0$. These are
\begin{align}
    \dot{\ab} = \Omegab\,\ab \qquad \text{and}\qquad  
    \dot{\eb} = \Lb_g\,\eb 
\end{align}
Figure \ref{fig:2} shows this difference for a remodelled fibre and a remodelled line element, which at time $t=\overline{t}$ coincides. We remark that since $\Rb$ is an orthogonal tensor the length of $\aref$ is unchanged whereas $\eb$ can be stretched (with a stretching rate $\Db_g\eb$).

Finally, since $\arem=\Rb\aref$ and $\dot\arem=\Omegab\aref$, the time rate of the remodeled orientation tensor $\Arem$ is
\begin{equation}\label{dotA}
\dot\Arem=\dot\arem\otimes\arem+ \arem\otimes\dot\arem=[\Omegab,\Arem]\,,
\end{equation}
where we have made use of the commutator operator $[\cdot,\cdot]:\Lin\times\Lin\to\Skw$  such that $
[\Ab,\Bb]=\Ab\Bb-\Bb\Ab\,,\quad\forall\Ab,\Bb\in\Lin\,$.

%
%%
%For later use, it is convenient to calculate how velocity fields transform under the invariance requirement~\eqref{F+}. In particular, when $\Fe^+=\Fe\Qb^T$ and $\Fv^+=\Qb\Fv$, one has
%%
%%\begin{equation}\label{CeAv}
%%\dot\Cb_e^+=\Qb\dot\Cb_e\Qb^T + [\Omega,\Cb_e^+]\quad\textrm{and}\quad
%%\dot\Arem_v^+=\Qb\dot\Arem_v\Qb^T + [\Omega,\Arem_v^+]\,,
%%\end{equation}
%%%
%%and 
%%
%\begin{equation}\label{LvDv}
%\Lb_v^+=\Qb\Lb_v\Qb^T+\Omegab\quad\textrm{and}\quad
%\Db_v^+=\Qb\Db_v\Qb^T\,, 
%\end{equation}
%%
%with $\Omegab=\dot{\Qb}\Qb^T\in\Skw$. These two expressions will be used in the next section to introduce suitable simplification in the constitutive functions.
%
\section{Balance equations}
\label{Beqn}
The principle of virtual working defines the weak balance equations of the model and, through the proper localization, allows us to introduce the standard local balance of forces and the new local balance equations of the torques working-conjugate of the fiber reorientation and of the couples working-conjugate of the matrix remodeling actions.\\
In doing so we call $\zb$ and $\mathbf s$ the forces per unit of (referential) volume and  area, $\Yb$ and $\Zb$ the external couple and torque per unit of (referential) volume. On the other hand, we assume that internal working is defined in terms of the (first) Piola--Kirchhoff stress tensor $\Sb$, the stress-couple $\Gb$ and the stress-torque $\Sigmab$. Both the  external and internal workings are continuous, linear, real-valued functional on the space of virtual rates $(\dot{\tilde{p}},\widetilde\Lb_g,\widetilde{\Omegab})$, given by
\begin{equation}
\Wc_e(\tilde{\dot p},\tilde\Lb_g,\tilde{\Omegab})=\int_\Bc(\zb\cdot\tilde{\dot p} +  \Yb\cdot\tilde\Lb_g+\Zb\cdot\tilde{\Omegab}) + \int_{\partial\Bc}\mathbf s\cdot\tilde{\dot p}
\end{equation}
for the external working, and
\begin{equation}
\Wc_i(\tilde{\dot p},\tilde\Lb_g,\tilde{\Omegab})=\int_{\Bc}(\Sb\cdot\nabla\tilde{\dot p}  + \Gb\cdot\tilde\Lb_g +\Sigmab\cdot\tilde{\Omegab})\,,
\label{InternalWorking}
\end{equation}
for the internal one. 

The principle of virtual working states that for any given subregion $\Pc\subset\Bc_r$ of the body, the external and internal workings must be equal for all virtual velocities $(\wb,\tilde{\Lb}_g,\tilde{\Omegab})\in\Vc\times\Lin\times\Skw$. Therefore, through a standard localization argument, the following strong form of the balance equations can be derived together with the corresponding boundary conditions\footnote{Boundary conditions are only referred to the standard balance of forces as the internal working for stress-couple and stress torque is of  zero order, that is, no internal actions expend working on the gradient of $\tilde{\Lb}_g$ and $\tilde{\Omegab}$.}:
\begin{equation}\label{be}
\begin{aligned}
&\Div\Sb + \zb =\0\qquad \textrm{in}\,\,\Bc\,,\\
&\ub=\hat\ub\,\,\,\textrm{in}\,\,\partial_u\Bc\\ &\Sb\,\mb=\mathbf{s}\,\,\,\textrm{on}\,\,\partial_t\Bc\,,
\end{aligned}
\end{equation}
and
\begin{equation}
\label{microbe}
\Gb=\Yb\qquad\textrm{and}\qquad
\Sigmab=\Zb\,\,\,
\end{equation}
with $\partial_u\Bc$ and $\partial_t\Bc$ parts of the boundary $\partial\Bc$ where displacements and tractions are prescribed and $\mb$ the unit normal to $\partial_t\Bc$.
The former equation \eqref{be} is the standard balance equation of the linear momentum expressed in terms of the first Piola--Kirchhoff stress tensor, whereas the latter \eqref{microbe} is the balance equations of the stress-couples and stress-torques arising from our zero-th order theory.

%
%%%%%%%%%%%%%%%%%%%%%%%%%%%%%%%%%%%%%%%%%%%%%%%%%%%%%%%%%%%%%%%%%%%%%%%
%\subsection{Power, elastic energy and dissipation}
%%%%%%%%%%%%%%%%%%%%%%%%%%%%%%%%%%%%%%%%%%%%%%%%%%%%%%%%%%%%%%%%%%%%%%
%
%Balance equations  reduce the outer working and identifies the power expended during the evolution of the body as  $\Wc_i(\dot p,\Lb_g,\Omegab)$.
%
%It is related to the elastic energy $\varphi$ and the dissipation function $\delta$ by the  dissipation inequality, which has to be satisfied by any thermodynamically consistent constitutive prescription of the stress $\Sb$, the stress-couple $\Gb$ and the stress-torque $\Sigmab$.\\
%
 
%
\section{Structural frame-indifferent constitutive prescriptions}
\label{sec:SFI}

The multiplicative decomposition of the deformation gradient \eqref{kronerlee} causes the natural state to be not unique since both $(\Fb_g,\Rb)$ and $(\Qb\Fb_g,\Qb\Rb)$ gives the same macroscopic deformation as Eqs.~\eqref{md1}-\eqref{md2} have evidenced. To this respect, the Authors of \cite{Dicarlo:2002} wrote:  \textit{...there is no reason why the stress response from $\Qb\Fb_g$ should be $\Qb-$related to the one from $\Fb_g$. This does happen, however, if the body element is isotropic and its relaxed state undistorted \cite{Truesdell:1991}} (see also \cite{Davini:2001}).

The $\Qb-$ relation cited in \cite{Dicarlo:2002} is the main idea behind the so called principle of structural frame-indifference (SFI), first formulated in \cite{Green:1971}. According to Green and Naghdi, the non-uniqueness of the natural state must not influence the constitutive response of the continuum whether or not the material is isotropic. An immediate consequence is that all constitutive functions must be insensitive to the transformation laws \eqref{transformation}. %, since they produce the same macroscopic effect, whether the body is isotropic or not. %of the , that states that \\ 

Differently from what asserted in \cite{Dicarlo:2002}, we require SFI to holds and we examine the consequences of such a choice on the constitutive functions of the model. In particular, our modelling framework requires the introduction of two constitutive functions, the strain energy density $\varphi$ and the dissipation density $\delta$, both per unit of mass, that allow us to completely characterize the material response. Since the material is transversely isotropic, we assume $\varphi$ to depend on the left Cauchy-Green strain tensor $\Cb_e$ and on the remodelled structural tensor $\Arem=\Rb\Aref\Rb^T$, i.e., $\varphi=\varphi(\Cb_e,\Arem)$.
%such that $\varphi=\varrho_r\hat\varphi$, with $\varrho$ the reference density. It corresponds to a transversely isotropic material with transverse isotropy axis evolving in the natural state as $\arem=\Rb\aref$ does. 
Likewise, we require the dissipation function $\delta$ to depend on the inelastic rates $\Lb_g$ and $\Omegab$, such that $\delta=\delta(\Lb_g,\Omegab)$. %; our choice identifies the processes which deliver dissipative contribution.
With these assumptions both $\delta$ and $\varphi$ are frame indifferent, i.e, the theory is objective. In addition, SFI imposes some restrictions on the constitutive functions that have to be carefully assessed.

The transformation laws \eqref{transformation} make the argument of $\varphi(\Cb_e,\Arem)$ changing as
\[
    \Cb_e\mapsto \Qb\Cb_e\Qb^T,\qquad \Arem \mapsto \Qb\Arem\Qb^T,
\]
%with $\Qb\in\Orth^{+}$. 
thus the requirement that the strain energy density is structurally frame-indifferent gives
\begin{equation}\label{Qe}
    \varphi(\Cb_e,\Ab)=\varphi(\Qb\Cb_e\Qb^T,\Qb\Arem\Qb^T),
\end{equation}
for any $\Qb\in\Orth^+$. Eq.~\eqref{Qe} is indeed satisfied for every rotation $\Qb$ if and only if $\varphi$ is a isotropic function of the two tensors $\Cb_e$ and $\Ab$. This allows the energy density to be expressed in terms of the invariants of the two tensors (see \cite{Ciambella:2021}).

For what concerns the dissipation function $\delta(\Lb_g,\Omegab)$, its arguments change as
\begin{equation}
    \Lb_g\mapsto\dot{\Qb}\Qb^T+\Qb\Lb_g\Qb^T,\qquad 
    \Omegab\mapsto \dot{\Qb}\Qb^T+\Qb\Omegab\Qb^T,
\label{InelasticRateTransf1}
\end{equation}
or in terms of the symmetric and the skew-symmetric parts of the inelastic strain rate
\begin{equation}
\Db_g\mapsto \Qb\Db\Qb^T,\qquad
\Wb_g\mapsto \dot{\Qb}\Qb^T+\Qb\Wb_g\Qb^T.
\label{InelasticRateTransf2}
\end{equation}
By separating the inelastic stretch rate and the inelastic spin rate in the argument of $\delta$, the principle of SFI requires that 
\begin{equation}\label{Qd}
    \delta(\Db_g,\Wb_g,\Omegab)=\delta(\Qb\Db_g\Qb^T,\dot{\Qb}\Qb^T + \Qb\Wb_g\Qb^T,\dot{\Qb}\Qb^T+\Qb\Omegab\Qb^T)\,,
\end{equation}
for any $\Qb\in\Orth^+$ and for any $\dot{\Qb}\Qb^T\in\Skw$. %Equation~\eqref{Qd} shows that the inelastic spin $\Wb_g$ and the reorientation spin $\Omegab$ share the same indeterminacy due to the arbitrariness of $\dot{\Qb}\Qb^T$. 
Due to the arbitrariness of $\Qb$ and $\dot{\Qb}\Qb^T$, one can choose $\Qb=\Ib$ and $\dot{\Qb}\Qb^T=-\Wb_g$ and write
\begin{equation}\label{Qd1}
    \delta(\Db_g,\Wb_g,\Omegab)=\hat\delta(\Db_g,-\Wb_g + \Wb_g,-\Wb_g+\Omegab)=\delta(\Db_g,\0,\Omegab-\Wb_g)\,,
\end{equation}
meaning that the dissipation function can only depend on the inelastic stretch rate $\Db_g$ and on the difference between the reorientation spin rate and the inelastic spin rate $\Omegab-\Wb_g$. 
Therefore, with a slight abuse of notation, we will drop the dependence on $\Omegab+\Wb_g$ from $\delta$ to write the following structurally frame-indifferent form of the dissipation function
\begin{equation}\label{Qd2}
    \delta=\delta(\Db_g,\Omegab-\Wb_g)\,.
\end{equation}
Let us note that if dependence of $\delta$ on the evolving material structure $\Arem$ is incorporated in the function, previous results still hold true.
%that is independent of  $\Omegab+\Wb_g$. %does not affect the dissipation, whichever value it gets. 
%It is worth noting that structural frame indifference  transforms the relative spin as
%
%\begin{equation}
%    \Omegab-\Wb_g\mapsto \Qb(\Omegab-\Wb_g)\Qb^T\,;
%\end{equation} 
%
%on the contrary, $\Omegab+\Wb_g$ is indetrminate:
%
%\begin{equation}
%    \Omegab+\Wb_g\mapsto 2\dot{\Qb}\Qb^T + \Qb(\Omegab+\Wb_g)\Qb^T\,.
%\end{equation} 
%
%
\section{Thermodynamic consistency of the constitutive equations}
\label{sec:tc}
The second principle of thermodynamics in absence of temperature effects, is reduced to the local form of the dissipation inequality that prescribes the time rate of the internal elastic energy be less than or equal to the external working, or in other term that the dissipation function is positive, i.e., $\delta=\Wc_{e}-\frac{d}{dt}\varrho_r\varphi\geq 0$, with $\varrho_r$ the mass density per unit of volume in the reference configuration. Due to the principle of virtual working, the dissipation inequality can be equivalently written in terms of the internal working. \\
We point out that the thermodynamic restrictions are imposed on the space of \emph{admissible velocity fields}; an immediate consequence is that reactive part of stresses, stress-couples or stress-torques do not enter the inequality as they must expend null working on the admissible velocity fields. We hence write
\begin{eqnarray}
\delta=&&\hat\Sb\cdot\dot\Fb + \sym\hat\Gb\cdot\Db_g + \frac{\hat\Sigmab-\skw\hat\Gb}{2}\cdot(\Omegab-\Wb_g) \nonumber\\
&+&
\frac{\hat\Sigmab+\skw\hat\Gb}{2}\cdot(\Omegab+\Wb_g)-\varrho_r\dot\varphi \geq 0\,.\label{DI}
\end{eqnarray}
where we have indicated with a superimposed hat $\hat{\,}$ the constitutively assigned parts of the Piola-Kirchhoff stress $\hat{\Sb}$, of the stress-couple $\hat{\Gb}$ and of the stress-torque $\hat{\Sigmab}$. The internal working \eqref{InternalWorking} was rewritten in the form \eqref{DI} to highlight the role of the relative spin $\Omegab-\Wb_g$. 

The time derivative of the strain energy density $\dot{\varphi}={\partial\varphi}/{\partial \Cb_e}\cdot \dot{\Cb_e} +{\partial\varphi}/{\partial \Ab}\cdot \dot\Ab$ can be rewritten in view of Eqs.~\eqref{rates1} and \eqref{dotA} as
\begin{equation}
\varrho_r\dot{\varphi}=\Fb_e\hat\Sb_e\Fb_e^T\cdot\Db -\sym(\Cb_e\hat\Sb_e)\cdot\Db_g + \dfrac{1}{2}[\Cb_e, \hat\Sb_e]\cdot(\Omegab-\Wb_g)\,,
\label{tre'}
\end{equation}
where we have defined the symmetric relaxed Piola-Kirchhoff stress  $\hat\Sb_e=2\varrho_r\partial\hat\varphi/\partial\Cb_e$ and we have used the identity 
 \begin{equation}
     \varrho_r[\dfrac{\partial\hat\varphi}{\partial \Ab},\Ab]=\frac{1}{2}[\Cb_e,\hat\Sb_e]\,,
 \end{equation}
proved true for a transversely isotropic material in \cite{Ciambella:2019}. It is worth remarking that $\Mb=\Cb_e\hat\Sb_e$ is the so-called \emph{Mandel stress}, for which $\skw\lbrace\Mb \rbrace = \frac 12 [\Cb_e,\hat\Sb_e]$ and $\sym\Mb = \sym(\Cb_e\hat\Sb_e)$. 

With \eqref{tre'} on hands, we rewrite the dissipation inequality \eqref{DI} and we introduce the suitable  constitutive equations of $\hat\Sb$, $
\hat\Gb$ and $\hat\Sigmab$ for it to holds for any velocity fields. One has
\begin{align}
\delta = & \big(\hat\Sb \Fb^T-\Fb_e\hat\Sb_e\Fb_e^T\big)\cdot \Db
+\big(\sym\hat\Gb +\sym(\Cb_e\hat\Sb_e)\big)\cdot\Db_g\label{DI1}\\[0.2cm]
&+\big(\frac{\hat\Sigmab-\skw\hat\Gb}{2} -\dfrac{1}{2}[\Cb_e, \hat\Sb_e] \big)\cdot(\Omegab-\Wb_g)+\frac{\hat\Sigmab+\skw\hat\Gb}{2}\cdot(\Omegab+\Wb_g)\geq 0\nonumber\,,
 \end{align}
 that must hold true for any admissible velocities $(\Db,\Db_g,\Wb_g,\Omegab)\in\Sym\times\Sym\times\Skw\times\Skw$. 
 
The principle of SFI applied to the dissipation function requires $\delta$ be dependent only on $\Db_g$ and $\Omegab-\Wb_g$, (see Eq.~\eqref{Qd2}). Therefore the following constitutive choices immediately follows
 \begin{equation}\label{stresses}
 \hat\Sb \Fb^T=\Fb_e\hat\Sb_e\Fb_e^T\qquad\textrm{and}\qquad
 \hat\Sigmab+\skw\hat\Gb=\0\,.
 \end{equation}
 Equation \eqref{stresses}$_1$ determines the constitutive equation of the Piola-Kirchhoff stress $\hat\Sb$, whereas equation \eqref{stresses}$_2$ prescribes that the sum of the constitutively determined components of $\hat\Sigmab$ and $\skw\hat\Gb$ is null. %We remark that the reactive parts of the stresses do not enter the dissipation inequality since they must expend null working on the velocity fields. 
 With the assumptions \eqref{stresses}, the dissipation inequality becomes %the following reduced form
 \[
 \delta = \big(\sym\hat\Gb +\sym(\Cb_e\hat\Sb_e)\big)\cdot\Db_g +\big(\frac{\hat\Sigmab-\skw\hat\Gb}{2} -\dfrac{1}{2}[\Cb_e,\hat\Sb_e] \big)\cdot(\Omegab-\Wb_g)\geq 0
 \]
 that suggests the following thermodynamically consistent constitutive prescriptions:
 \begin{equation}\label{stresses2}
 \begin{aligned}
 &\sym\hat\Gb=-\sym(\Cb_e\hat\Sb_e)+\Do\Db_g\,,\\
 &\hat\Sigmab-\skw\hat\Gb=[\Cb_e, \hat{\Sb}_e]+2\Ko(\Omegab-\Wb_g)\,,
 \end{aligned}
 \end{equation}
 with $\mathbb{D}$ and $\mathbb{K}$ fourth-order positively definite tensors, that guarantee $\delta$ be a positive definite quadratic form of the strain rates. It is pointed out that choices more complex than \eqref{stresses2} still compatible with thermodynamics are indeed possible (see for instance \cite{Ciambella:2021}), yet the constitutive prescriptions \eqref{stresses2} allows us to highlight the main features of our theory and is used for the sake of simpleness.
\section{Evolution equations of the inelastic processes}
\label{ev:eqn}
Let us start by splitting the balance equation~\eqref{microbe}$_2$ into its symmetric and skew-symmetric parts
\begin{equation}\label{be2}
    \sym\hat\Gb+\sym\tilde\Gb=\sym\Yb\quad\textrm{and}\quad\skw\hat\Gb+\skw\tilde\Gb=\skw\Yb\,.
\end{equation}
where we have explicitly indicated the constitutively determined part of stress-couple $\hat\Gb$, and in its reactive part, $\tilde{\Gb}$. When the constitutive equation \eqref{stresses2}$_1$ for $\sym\Gb$ is used within equation \eqref{be2}$_1$, we get the following flow rule
\begin{equation}\label{una11}
\Do\Db_g=\sym\Yb -\sym\tilde\Gb +\sym(\Cb_e\hat\Sb_e)
\,,    
\end{equation}
which drives the evolution of $\Db_g$ in terms of the external sources $\sym\Yb$ and of the symmetric part of the tensor $\Cb_e\hat\Sb_e$, i.e., the symmetric part of the Mendel stress. Analogously, subtraction of Eqs.\eqref{microbe} from \eqref{be2}$_2$ and substitution into %
%
%\begin{equation}\label{besplit}
%\Sigmab-\skw\Gb = \Zb-\skw\Yb\quad\textrm{and}\quad
%\Sigmab+\skw\Gb = \Zb+\skw\Yb\,.
%\end{equation}
%
%When 
the constitutive equation \eqref{stresses2}$_2$ give 
\begin{equation}\label{due2}
2\,\Ko(\Omegab-\Wb_g)=\Zb-\skw\Yb -\tilde{\Sigmab} + \skw\tilde\Gb - [\Cb_e,\hat\Sb_e]\,,   
\end{equation}
which drives the evolution of the spin difference $(\Omegab-\Wb_g)$ in terms of the external sources $\Zb-\skw\Yb$ and of the skew-symmetric part of the Mendel stress tensor $\Cb_e\hat{\Sb}_e$. In Eq.~\eqref{due2} a superimposed tilde $\tilde{\,}$ was again used to indicate the reactive components of the stress torques $\tilde{\Sigmab}$ and of the stress couples $\skw\tilde\Gb$.\\ 
On using the relationship \eqref{stresses}$_2$ between the constitutively determined parts of $\Sigmab$ and $\Gb$, the sum of the balance equation \eqref{microbe} with the second of \eqref{be2} gives
\begin{equation}\label{besplit}
%\Sigmab-\skw\Gb = \Zb-\skw\Yb\quad\textrm{and}\quad
\tilde{\Sigmab}+\skw\tilde{\Gb} = \Zb+\skw\Yb\,,
\end{equation}
In order to satisfy this balance equation, two different choices are indeed possible: one can restrict the range of allowable external actions in a way that $\Zb+\skw\Yb=\0$ and so $\tilde\Sigmab+\skw\tilde\Gb = \0$, or one admits the existence of reactive components in \eqref{besplit} by introducing the kinematical constraint
%equation, together with \eqref{stresses}$_2$ show that, to keep the theory consistent with SFI, external sources $\Zb+\skw\Yb$ have to be null within the theory or we must provide any not constitutively determined $\tilde{\Sigmab}+\skw\tilde{\Gb}$ which can balance those external sources: 
%
%\begin{equation}\label{extra}
%    \tilde{\Sigmab}+\skw\tilde{\Gb} = \Zb+\skw\Yb\,.
%\end{equation}
%
%Being $\Omegab+\Wb_g$ indeterminate, we suggest to take  
%
\begin{equation}\label{constr}
    \Omegab+\Wb_g\equiv\0\,.
\end{equation}
In this latter case, the skew symmetric components of the external actions are balanced by the reactive components in \eqref{besplit}.  In this paper, coherently with the development in \cite{Ciambella:2021}, we will follow this second path by restricting the space of allowable veloctiy fields with the constraint \eqref{constr}. The consequence os such a choice are now analyzed in details.

We first note that the reactive components of the internal actions must expend null working on the admissibile fields. Hence we write
%all along the inelastic processes, that is, to introduce an internal constraint on the space of admissible velocity fields. Thus, it has to be maintained by the proper reactive actions. By indicating the reactive stresses with a $\tilde{\,}$, they must expend a null working so that
%
\begin{equation}\label{powerR}
\tilde{\Sb}\cdot\dot\Fb + \sym\tilde{\Gb}\cdot\Db_g+\frac{\tilde{\Sigmab}+\skw\tilde{\Gb}}{2}\cdot(\Omegab+\Wb_g) +\frac{\tilde{\Sigmab}-\skw\tilde{\Gb}}{2}\cdot(\Omegab-\Wb_g) = 0\,,
\end{equation}
on any admissible motion satisfying~\eqref{constr}. Since no constraint is imposed on $\dot{\Fb}$, $\Db_g$ and $\Omegab-\Wb_g$, the corresponding reactive stresses vanish $\tilde{\Sb}=\0$, $\sym\tilde{\Gb}=\0$,  $\tilde{\Sigmab}-\skw\tilde{\Gb} =\0$, whereas, in general,
%
%\begin{equation}
$\tilde{\Sigmab}+\skw\tilde{\Gb}$   
%\end{equation}
%
 is not null in view of the balance equation \eqref{besplit}. With this on hand, we can rewrite the equations governing the evolution problem in their pure forms:
 \begin{align}
&\Do\Db_g=\sym\Yb +\sym(\Cb_e\hat\Sb_e) \label{Evol1}\\
&2\,\Ko(\Omegab-\Wb_g)=\Zb-\skw\Yb - [\Cb_e,\hat\Sb_e]\,,  \label{Evol2} 
 \end{align}
that together with the macroscopic balance equation
%Finally, the balance of forces and the constitutive prescription for $\hat\Sb$ furnishes, as $\tilde{\Sb}=\0$, the pure equations
%
\begin{equation}\label{tre3}
    \dvg\hat\Sb + \zb =\0\quad\textrm{with}\quad\hat\Sb\Fb^T=\Fb_e\hat\Sb_e\Fb_e^T\,,
\end{equation}
allow us to solve the equilibrium problem once the proper boundary and initial conditions are specified.
%which must be solved with the boundary conditions on displacements or tractions: $\ub=\hat\ub$ on $\partial_u\Bc$ and $\Sb\,\mb=\mathbf{s}$ on $\partial_t\Bc$.\\
%
Equations \eqref{una11}, \eqref{due2} and \eqref{tre3} together with the constraint equation \eqref{constr}  determine the evolution of the 12 unknowns of our theory $p$, $\Fb_g$ and $\Rb$.
\subsection{The reduced evolution problems}
The evolution of the inelastic variables of the theory requires the solution of Eqs.~\eqref{Evol1} and \eqref{Evol2} once the proper initial conditions are specified. In this subsection we will examine possible applications of our theory to prototypical examples from the literature.

We start by considering a anisotropic material constituted by a viscous matrix reinforced with stiff fibers. As is customary in the literature, the external actions acting on the matrix inelastic variables are considered to be null, that is  $\sym\Yb=\skw\Yb=\0$ (see \cite{Reese:1998,Ciambella:2021}). On the other hand, fibre reorientation may be driven by external sources, thus we assume $\Zb\not=\0$ (see for instance \cite{Ciambella2:2019} for fibre reorientation driven by the magnetic field); we also assume  null bulk forces $\zb$. In this circumstance Eqs.~\eqref{Evol1}, \eqref{Evol2} and \eqref{constr} give 
\begin{equation}\label{una112}
\eta_d\Db_g=\sym(\Cb_e\hat\Sb_e)\quad\textrm{and}\quad    
\eta_r\Omegab=\Zb- \frac{1}{2}[\Cb_e,\hat\Sb_e]\quad\textrm{and}\quad\Wb_g=-\Omegab\,,
\end{equation}
to be solved with the initial conditions $\Fb_g(X,0)=\Ib$ and $\Rb(X,0)=\Ib$. In writing Eqs. \eqref{una112}$_{1,2}$, we have assumed that the remodelling tensors are isotropic, that is $\Do=\eta_d\Io$ and $\Ko=4\eta_r\Io$, with $\eta_d$ and $\eta_r$ the matrix and fiber viscosity, respectively.

The twelve equations in \eqref{una112} are coupled but can be numerically solved together with the macroscopic balance Eq.~\eqref{tre3} to get the twelve unknown fields in $\Fb_g$, $\Rb$ and $p$ governing the time evolution of the problem. It is worth noting that the system does not admit an equilibrium solution, in fact the application of the external field $\Zb$ steers the direction of the fibres within the viscous matrix which passively grows and influences fibre reorientation. In such a case, an external source $\sym\Yb$ would be needed to maintain the equilibrium solution determined by the equations
\begin{equation}
    \sym\lbrace\Cb_e\hat\Sb_e\rbrace=\sym\Yb\quad\textrm{and}\quad
    \frac{1}{2}[\Cb_e,\hat\Sb_e]=\Zb\,.
\end{equation}
\\
When no external actions are imposed, i.e., $\sym\Yb=\0$ and $\Zb=\0$, the only equilibrium solution of \eqref{una112} corresponds to the natural state at which $\Cb_e=\Ib$ and both $\sym\lbrace \Cb_e\hat\Sb_e\rbrace$ and $[\Cb_e,\hat\Sb_e]$ vanish. However, in this situation $\Rb$ is indeterminate, since the Mandel stress is zero in the natural state whether rotation $\Rb$ is considered. This apparent limit of the theory can be overcome by suitably prescribing a different dependence of the elastic energy on the rotation $\Rb$ or relaxing the internal constraint, as discussed in Sec.~\ref{Cefd}.

Another interesting example that the our theory can encompass is the growth problem of a continuum in which the fibre cannot reorient independently of the matrix. In such a case the following additional constraint on the rotation matrix has to be enforced 
\begin{equation}\label{RI}
    \Rb=\Ib\quad\textrm{or equivalently}\quad\Omegab=\0\,,
\end{equation}
that implies $\Arem\equiv\Aref$, i.e., fibres do not rotate from the reference configuration to the natural state. Equation~\eqref{RI} is indeed a constraint acting on the field $\Rb$, thus limiting the evolution of the state variables of the problem; therefore the proper reactive actions appear in the equations. In particular, the evolution equations of the dissipative process under the constraint \eqref{RI} are
\begin{equation}\label{una113}
m_d\Db_g=\sym\Yb+\sym\lbrace \Cb_e\hat\Sb_e\rbrace\quad\textrm{and}\quad\Wb_g=\0\,.
\end{equation}
The remaining balance equations allow the reactive stresses to be determined from the external actions
\begin{equation}
    \tilde\Sigmab+\skw\tilde\Gb = \Zb+\skw\Yb\quad\textrm{and}\quad
    \tilde\Sigmab-\skw\tilde\Gb = \Zb-\skw\Yb\,.
\end{equation}
A typical application of Eq.~\eqref{una113} is the growth of anisotropic tissues where the reinforcing fibre structure does not evolve from the reference configuration to the natural state, and the external field $\sym\Yb$ is used to bring into the modelling the effects of external stimuli \cite{Goriely:2017}.\\
We finally note that by assuming $\Fb_g=\Ib$, that is, $\Fb_e=\Fb$, the present formulation does not allow us to recover the elastic reorientation theory presented in \cite{Ciambella:2019}. In fact, by assuming $\Fb_g\equiv \0$, we would have  $\Lb_g=\0$,  $\Db_g=\Wb_g=\0$ and, hence $\Omegab=\0$. In this case, external sources, if any, determine the reactive stress-couples and stress-torques whereas Eq.~\eqref{tre3} determines the solution of the purely elastic anisotropic problem. 

\section{Asymptotic approximations}
\label{aa}
In order to further exploit the peculiarities of the proposed theory, we investigate the solutions of the evolution equations \eqref{una112} in the limit of slow or fast applied deformations, when no external actions are applied. We rewrite Eqs.~\eqref{una112} in the following form
\begin{equation}\label{una114}
\mu\,\tau_d\Db_g=\sym\lbrace\Cb_e\hat\Sb_e\rbrace\quad\textrm{and}\quad    
\mu\,\tau_r\Wb_g= \frac 12 \big[\Cb_e,\hat\Sb_e\big]\quad\textrm{and}\qquad\Wb_g+\Omegab=\0\,,
\end{equation}
where we have made explicitly the dependence on the characteristic times $\tau_d$ and $\tau_r$ defined from $\eta_d$ and $\eta_r$ as
$\tau_d=\eta_d/\mu$, $\tau_r=\eta_r/\mu$, where $\mu$ is the shear modulus of the matrix. We also introduce the characteristic deformation time defined by 
\[
\tau_c^{-1} = \vert \Db \vert,\qquad \text{such that}\quad \overline{\Db}=\tau_c\,\Db,\;\overline{\Db}_g = \tau_c \,\Db_g
\]
to obtain the following dimensionless evolution equations
\begin{equation}\label{Dimless}
\mu\,\frac{\tau_d}{\tau_c}\overline{\Db}_g=\sym\lbrace\Cb_e\hat\Sb_e\rbrace\quad\textrm{and}\quad    
\mu\,\frac{\tau_r}{\tau_c}\overline{\Wb}_g= \frac 12 \big[\Cb_e,\hat\Sb_e\big]\quad\textrm{and}\quad\overline{\Wb}_g+\overline{\Omegab}=\0\,,
\end{equation}
In the next part of this section two evolution regimes will be considered, one in which the characteristic deformation time is much longer that the characteristic times of the inelastic processes, that we call \emph{slow deformation}, and an opposite case in which the characteristic times of the deformation are much shorted than those driving the evolution and in this case we say that the deformation is \emph{fast}.

\emph{Slow deformations}. We first examine the case in which the applied deformation is slow by formally writing  that ${\text{max}\lbrace\tau_d,\tau_r\rbrace}/{\tau_c}\ll 1$, meaning that the matrix has relaxed around its natural configuration. We call $\varepsilon={\tau_d}/{\tau_c}$ such that $\varepsilon\ll 1$ and ${\tau_r}/{\tau_c} = \varepsilon\, {\tau_r}/{\tau_d}$, so that all variables can be expanded around the natural configuration in terms of the smallness parameter $\varepsilon$. One obtains
\begin{align}
    &\Fb_e = \Ib + \varepsilon\;\Fb_1\,,\\
    &\Fb_g = \big(\Ib-\varepsilon \Fb_1\big)\Fb + \text{o}(\varepsilon).
\end{align}
Accordingly
\begin{align}
&\Cb_e = \Ib + \varepsilon \big(\Fb_1+\Fb_1^T\big)+\text{o}(\varepsilon)\simeq \Ib +2\,\varepsilon \,\Eb_e ,\label{Ce}\\
&\overline{\Lb}_g = \overline{\Lb} - \varepsilon \big(\dot{\overline{\Fb}}_1+\big[\Fb_1,\overline{\Lb}\big]\big) + \text{o}(\varepsilon)
\end{align}
where the symbol $\simeq$ stands for first order approximation in $\varepsilon$. The strain tensor $\Eb_e$ in \eqref{Ce} is defined by $\Eb_e= \frac 12 \big(\Fb_1+\Fb_1^T\big)$. The constraint \eqref{Dimless}$_3$ gives 
\[
\overline{\Omegab} = -\overline{\Wb}_g \simeq -\overline{\Wb}\,.
\]
In addition the Mendel stress tensor becomes
\begin{equation}
\Cb_e \hat{\Sb}_e = \big(\Ib+2\varepsilon \Eb_e \big)\big(\hat{\Sb}_e(\Ib,\Rb)+ \varepsilon\mathbb{C}[\Eb_e] \big) + \text{o}(\varepsilon) \simeq \varepsilon \,\mathbb{C}[\Eb_e]\,,
\label{MendelSlowApproximation}
\end{equation}
where it was used the fact that the symmetric Piola stress tensor  vanishes in the natural state, i.e., $\hat{\Sb}_e(\Ib,\Rb)=\0$. The fourth order tensor $\mathbb{C}:=4\varrho_r{\partial^2 \varphi}/{\partial\Cb_e\partial\Cb_e}$ is the elasticity tensor evaluated around the natural state with symmetries dictated by $\Rb\Ab_0\Rb^T$. In this sense Eq.~\eqref{MendelSlowApproximation} shows that at the first order approximation of the Mendel stress coincides with the Cauchy stress of a transversely isotropic material. It is further noted that at the zero-th order the model predicts zero stress, which is a plausible result since the expansion has been carried out around the natural state.

\emph{Fast deformations}. When the characteristic deformation time $\tau_c$ is much smaller than the relaxation times which govern the evolution problem, the deformation is considered "fast". Formally we assume that  $\text{min}\lbrace \tau_d,\tau_r\rbrace \gg 1$, such that the $\iota = \tau_c/\tau_d \ll 1$ can be introduced. 
Accordingly, $\tau_c/\tau_r = \iota \tau_d/\tau_r$ and the following formal expansion can be considered
\begin{align}
    &\Fb_g = \Ib + \iota\, \Fb_1\,,\\
    &\Fb_e = \Fb\big(\Ib-\iota \,\Fb_1\big)+\text{o}(\iota)\,,\\
    &\Cb_e = \Cb-2 \iota \,\sym\lbrace\Cb\Fb_1 \rbrace+\text{o}(\iota)\,,
\end{align}
with the inelastic deformation rate given by
\begin{equation}
    \overline{\Lb}_g = \iota\,\dot{\overline{\Fb}}_1+\text{o}(\iota)\,.
\end{equation}
In such a regime, the Mendel stress tensor is evaluated as follows
\begin{align}
    \Cb_e\hat\Sb_e &= \big( \Cb-2\,\iota\,\sym\lbrace\Cb\Fb_1\rbrace\big)\big(\hat\Sb_e(\Cb,\Rb)-\iota\,\hat{\mathbb{C}}[\sym\lbrace\Cb\Fb_1\rbrace]\big)+\text{o}(\iota)\\
    &\simeq  \Cb\,\hat\Sb_e(\Cb,\Rb)-\iota\big(2\,\sym\lbrace \Cb\Fb_1\rbrace\hat\Sb_e(\Cb,\Rb)+\Cb\,\hat{\mathbb{C}}[\sym\lbrace\Cb\Fb_1\rbrace]\big)
\end{align}
where the elasticity tensor $\hat{\mathbb{C}}$ is evaluated around the current configuration at $\Cb_e=\Cb$ and $\Rb$. Therefore the model predicts at zero-th order a stress tensor $\Cb\,\hat\Sb_e(\Cb,\Rb)$ coincident with the one of a purely elastic anisotropic material with symmetries dictated by $\Rb\Aref\Rb^T$. At the first order, the evolution problem becomes
\begin{equation}
   \iota\, \sym\lbrace\dot{\overline{\Fb}}_1\rbrace = {\iota}\,\sym\lbrace\Cb\,\hat{\Sb}_e(\Cb,\Rb) \rbrace,\quad \text{and}\quad
   \iota\, \skw\lbrace\dot{\overline{\Fb}}_1\rbrace = \frac{\iota}{2}\frac{\tau_d}{\tau_r}\,\big[\Cb,\hat{\Sb}_e(\Cb,\Rb)\big]\,,
\end{equation}
together with the constraint equation $\overline{\Omegab}+\skw\lbrace\dot{\overline{\Fb}}_1\rbrace$ that completely determine the evolution of the system.
\section{Conclusions and perspectives}
\label{Cefd}
We have introduced a modelling framework capable of describing the mechanical response of anisotropic soft materials undergoing large inelastic deformations. The main idea behind the model is that the matrix and the internal fibre structure have two different relaxed states: the matrix can be elastically unloaded to a zero stress state such that its relaxed state differs from the one of the fibre. This assumption has allowed us to partially decoupled the evolution of the fibre structure from the one imposed by the matrix, making the interaction between the fibre and the matrix non--affine.  %is allowed to relax, passively or actively, that is, without or with direct growth stimuli, and growth is modeled through the multiplicative decomposition of the deformation gradient. Fibers are allowed to change their orientation and it is assumed that fiber reorientation is not dragged by the inelastic process which affects matrix. This requirement is made precise through the introduction of specific constitutive recipes.\\

Within this framework, the consistency of the model with thermodynamics and with the principle of structural frame-indifference was carefully discussed and their implications on the flow rules of the inelastic processes analysed.\\
In particular, it was shown that to fully determine the relaxed state, one can introduce an additional kinematical constraint which links the inelastic spin rate, due to the evolution of the matrix, to the reorientation spin rate of the fibres. From one hand, the constraint equation, together with the flow rules naturally arising from the dissipation inequality, make the evolution problem of the 12 unknowns of the problem, the placement $p$, the inelastic deformation $\Fb_g$ and the rotation tensor $\Rb$, fully determined. On the other hand, the constraint limits the scenarios attainable by the model, granted the considered constitutive assumptions which are indeed shared by other Authors in the literature \cite{Haupt:2002,Rubin:2012}.\\
In our opinion, the limits of the theory can be possibly overcome by assuming different constitutive prescriptions for $\varphi$ and $\delta$ with a  stronger interactions between fibres and matrix, as is the case in which the reoriented fibers are dragged by the inelastic processes that remodel the matrix. Interestingly, another possibility would be weaken the kinematical constrain by an elastic-type interactions for the reactions $\tilde\Sigmab+\skw\tilde\Gb$:  $\Omegab+\Wb_g = \Mo(\tilde\Sigmab+\skw\tilde\Gb)= \Mo(\Zb+ \skw\Yb)$. This latter circumstate, which is beyond the scope of the present paper, paves the way for future development of the theory.
\section*{Acknowledgements}
PN  acknowledges the support of MIUR (Italian Minister for Education, Research, and University) through the project PRIN 2017 n. 2017KL4EF3. JC acknowledges the support of MIUR through the project PRIN 2017 n. 20177TTP3S. PN and JC  wish to acknowledge the support of the Italian National Group of Mathematical Physics (GNFM-INdAM).


%

\bibliographystyle{spphys}       % APS-like style for physics
\bibliography{mybibfile}   % name your BibTeX data base
%

%
%% Non-BibTeX users please use
%%\begin{thebibliography}{}
%%
%% and use \bibitem to create references. Consult the Instructions
%% for authors for reference list style.
%%
%%\bibitem{RefJ}
%\end{thebibliography}

\end{document}